\documentclass{emulateapj}
\usepackage{apjfonts}
\usepackage{epsfig}

\shorttitle{The shaping of the Hubble fork up to redshift 1}
\shortauthors{Pannella et al.}

\begin{document}


\title{The evolution of the mass function split by morphology up to redshift 1\\ in the FORS Deep and the GOODS--S Fields\altaffilmark{**}}
\altaffiltext{**}{Based on observations collected at ESO, Chile (ESO Programmes 63.O-0005, 64.O-0149, 64.O-0158, 64.O-0229, 64.P-0150, 65.O-0048, 65.O-0049, 66.A-0547, 68.A-0013, 69.A-0014 and LP168.A-0485), and on observations made with  HST/ACS (GO Proposals 9425 and 9502), obtained from the ESO/ST-ECF Science Archive Facility.}


\author{M. Pannella\altaffilmark{1}, U. Hopp\altaffilmark{1,2}, R.P. Saglia\altaffilmark{1}, R. Bender\altaffilmark{1,2}, N. Drory\altaffilmark{3}, M. Salvato\altaffilmark{1}, A. Gabasch\altaffilmark{1,2} and G. Feulner\altaffilmark{1,2}}
\altaffiltext{1}{Max-Planck-Institut f\"ur extraterrestrische Physik, Giessenbachstr., Postfach 1312, D-85741 Garching, Germany; maurilio,saglia,mara@mpe.mpg.de}
\altaffiltext{2}{Universit\"atssternwarte M\"unchen, Scheinerstr. 1, D-81673 M\"unchen, Germany; hopp,bender,gabasch,feulner@usm.uni-muenchen.de}
\altaffiltext{3}{Department of Astronomy, University of Texas at Austin, 1 University Station C1400, Austin, TX 78712; drory@astro.as.utexas.edu}



\begin{abstract}
We study the evolution of the stellar mass density for the separate families of bulge--dominated  and disk--dominated galaxies over the redshift range  $0.25\le z\le1.15$. We derive quantitative morphology for a statistically significant galaxy sample of 1645 objects selected from the FORS Deep and the GOODS--S Fields. We find that the morphological mix evolves monotonically with time: the higher the redshift, the more disk systems dominate the total mass content. At $z\sim1$, massive objects (M$_*\ge7\times10^{10}$M$_\odot$) host about half of the mass contained in objects of similar mass in the local universe. The contribution from early and late type galaxies to the mass budget at $z\sim1$ is nearly equal. We show that {\it in situ} star formation is not sufficient to explain the changing mass budget. Moreover we find that the star formation rate per unit stellar mass of massive galaxies increases with redshift only for the intermediate and early morphological types, while it stays nearly constant for late-type objects. This suggests that merging and/or frequent accretion of small mass objects has a key role in the shaping of the Hubble sequence as we observe it now, and also in decreasing the star formation activity of the bulge--dominated descendants of massive disk galaxies.
\end{abstract}



\keywords{cosmology: observations --- galaxies: morphology --- galaxies: formation --- galaxies: evolution --- galaxies: mass function --- surveys}


\section{Introduction}
\label{sec_introduction}

\citet{lilly1996}, and soon after \citet{madau1996}, produced the first estimates of the cosmic star formation history. Since then, the growing number of deep extragalactic surveys has allowed galaxy evolution to be tackled in more and more detail, to higher and higher redshifts \citep[e.g.][]{giavalisco2004, gabasch:sfr, bouw2004,juneau2005}. All these studies point towards a substantial amount of star formation at early cosmic epochs. 

More recently, a number of studies, mainly based on NIR selected surveys like  2MASS \citep{cole2001}, MUNICS \citep{drory2004a}, K20 \citep{fontana2004}, FIRES \citep{rudnick2003}, HDF \citep{dickinson2003} and GOODS+FDF \citep{drory2005}, have been able to measure directly the stellar mass density up to high redshifts. The two independent approaches are in a remarkably good agreement, and suggest that about half of the present--day stars was already in place at z $\approx$ 1, when the Universe was half of its present age.

The assembly of the stellar mass through cosmic time is a crucial test for models of galaxy formation and evolution \citep{KC1998}. Such models aim at linking the hierarchical growth of dark matter structures to the observed galaxy properties, by means of simplified prescriptions for the formation of baryonic systems in dark matter haloes \citep[e.g.][]{cole2000,gd2004}. 

An effective way to put sensitive constraints on the models is to understand where the stars were located, specifically what was the morphology of their host galaxies, at different look-back times. In fact, this permits to directly probe how galaxies assembled their stars, and how their morphology (i.e., at least with some approximation, their dynamical status) evolves. 
Since merging is driving both mass assembly and dynamical evolution in a hierarchical scenario, these studies offer a direct probe of the role of this process in galaxy evolution. 

Because of the lack of sufficient angular resolution, galaxies have been often classified, both at low and high redshift, as early or late types based on their broad band colors \citep{baldry2004,fontana2004}. However this kind of approach suffers, especially at high redshift, from the obvious drawback that a disk galaxy populated by an old stellar population would be classified as an early--type object (and vice versa). \citet{bell2003} used the concentration parameter to discriminate between early and late type objects in a complete sample extracted from local surveys.  They found that in the local Universe the {\em transition mass}, i.e. the mass at which disks become dominant in the relative contribution to the total stellar mass function, is $\approx5\times10^{10}$M$_\odot$. 


In this work we rely on a deep, complete and automatically morphologically classified sample to study the contribution of galaxies of different morphologies to the redshift evolution of the stellar mass density. 
\citet{bundy2005}, and also \citet{brinchmanneellis2000}, have carried out a work qualitatively similar to the one we present here, but based on shallower samples and visual morphological classification.


This Letter is organized as follows: in \S \ref{sec_data} we discuss the dataset on which this work is based, in \S \ref{sec_sersic} we present the surface brightness profiles modeling, and in \S \ref{sec_mass} our results on the evolution of mass functions, mass densities and specific star formation rates (SSFR, i.e. star formation rate per unit stellar mass) split by morphological type. Finally, in \S \ref{sec_conclusions} we draw our conclusions.

We use AB magnitudes and adopt a cosmology with $\Omega_M$=0.3, $\Omega_\Lambda$=0.7, and H$_0$=70 kms$^{-1}$Mpc$^{-1}$.
\section{Ground-based data, photo--z and M$_*$/L ratios}
\label{sec_data}
This study is based on photometric catalogs derived for the FORS Deep Field \citep[FDF, ][]{fdf:data,gabasch:lf} and part of the GOODS--South Field (Salvato et al., in preparation). The two fields cover approximately the same area (39.8 arcmin$^2$ for FDF and 50.6 arcmin$^2$ for GOODS--S). The FDF photometric catalog covers the UBgRIzJKs broad bands, plus an intermediate band centered at 834nm. The I-selected catalog used here lists 5557 galaxies down to  $I_{lim}=26.8$. For the GOODS--S field, our K-band selected catalog  contains 3240 galaxies, and it consists of UBVRIJHK broad-band photometry \citep[][Vandame et al., in preparation]{arnouts2001,schirmer2003}. Number counts match the literature values down to $K_{lim}\approx 25$.

Photometric redshifts are derived using the method described in \citet{ben2001}. Their accuracy is 3\% and 5\% for FDF and GOODS--S respectively, with $\approx1\%$ outliers for both fields \citep[for more details see][]{gabasch:lf}. 

For all galaxies in both catalogs, the M/L ratios were estimated with a log--likelihood based SED fitting technique, using a library of SEDs built with the \citet{bruzualecharlot2003} code, assuming a \citet{salpeter1955} IMF. The procedure used is described in detail in \citet{drory2004b}. We use here the same mass catalog as in \citet{drory2005}, so we refer the reader to this paper for more details. 

\begin{figure}
\begin{center}
\includegraphics[angle=0,width=0.75\linewidth,bb = 19 145 580 700]{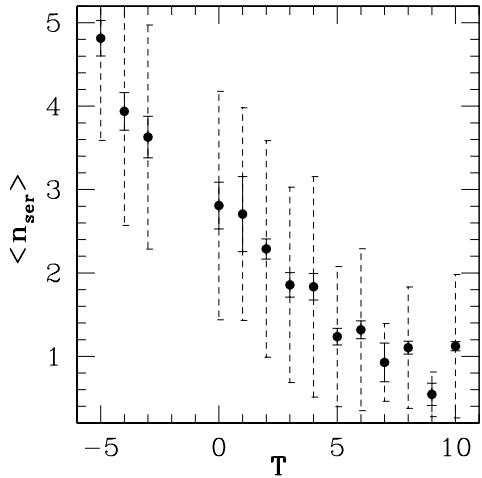}
\caption{Correlation between the Sersic index and the visual classification T value for the GOODS--S subsample. Points are mean values. Solid error bars are the errors on the means, dashed error bars show the rms values. \label{fig:morph_nt}}
\end{center}
\end{figure}

\section{HST imaging and the morphological analysis}
\label{sec_sersic}
Both the FDF and GOODS--S fields were imaged with the ACS camera on board the HST. The FDF was imaged in the broad--band F814W filter, with 4 WFC pointings of 40 minutes exposure each, reaching a 10$\sigma$ limit of 26 mag. The data reduction was performed with the standard {\sl CALACS\footnote {www.stsci.edu/hst/acs/analysis}} pipeline, and the combined final mosaic was produced with the multidrizzle package \citep{mutchler2003}. The GOODS--S field was imaged in four different filters (BVIz) as part of the ACS GOODS legacy \citep{giavalisco2004}; the 10$\sigma$ limit in the F775W band is 26.5 mag. 

We make use of the publicly available packages GIM2D \citep{simard1999} and GALFIT \citep{peng2002} to fit PSF convolved \citet{sersic1968} profiles to the two-dimensional surface brightness of each object, down to a limit of F814W=24 (FDF) and F775W=24.5 (GOODS--S). The PSFs used to convolve the profiles were obtained for each individual tile by stacking about 10 high S/N isolated stars. The results from the two different codes are in excellent agreement, thus confirming the robustness both of our modeling and of the choice of flux limits. 

More details on the reduction and the data quality assessment, as well as on the whole fitting procedure and relative validation, will be described in Pannella et al. (in preparation). In order to further check our morphological classification, visual morphology was performed by UH according to the \citet{devauc} classification scheme. A tight correlation is obtained between the average visual and automated classifications, parameterized by the morphological type $T$ and the Sersic index $n_{ser}$, respectively (see Fig.~\ref{fig:morph_nt}). Using the visual or automated classification does not significantly affect the results presented in this study. 

We split our sample, according to the Sersic index, in early--type (n $\ge$ 3.5, $\approx$ T $\le$ -3), intermediate (2 $\le$ n $<$ 3.5, $\approx$ -3 $<$ T $\le$ 2) and late--type (n $<$ 2, $\approx$ T $>$ 2). 
We evaluated the effect of morphological k-correction on this coarse classification scheme by also performing the surface brightness modeling on B and $z$--band images for a subsample of the GOODS-S galaxies. We found that our broad classification is robust for objects at $z \le 1.15$. Thus we restrict our analysis to $z \le 1.15$. 

Finally, we estimate the mass completeness of the two (FDF and GOODS-S) catalogs as the mass of a maximally old stellar population\footnote{We used a dust-free, passively evolving stellar population model, ignited by an instantaneous burst of sub solar metallicity at z = 10.} having, at every redshift, an observed magnitude equal to the catalog completeness magnitude \citep[e.g.][]{dickinson2003}. The final catalog contains 1645 galaxies. 
\begin{figure*}
\begin{center}
\includegraphics[angle=0,width=0.75\linewidth,bb = 19 145 589 386]{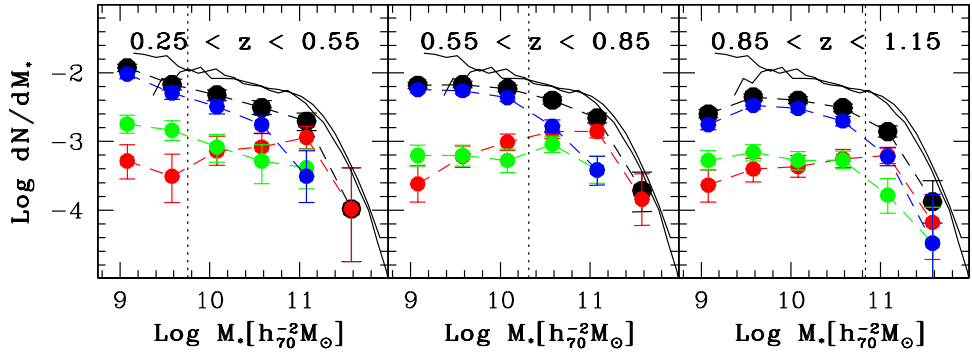}
\caption{The mass function split by morphological types (early in red, intermediate in green, and late in blue) in three redshift bins. Black symbols refer to total values. The vertical axis is in units of [h$_{70}^3$Mpc$^{-3}$dex$^{-1}$]. A vertical dotted line indicates the mass completeness limit in each redshift bin. Dashed colored lines are intended to guide the eye. Solid black lines show the local mass function determinations from \citet{cole2001} and  \citet{bell2003}. \label{fig:morph_MF}}
\end{center}
\end{figure*}
\section{The evolution of the morphological mass function, total mass density and SSFR}
\label{sec_mass}
We performed extensive Monte-Carlo simulations to take into account the effect of mass uncertainties ($\approx0.2$dex) on our results. Ten thousand simulations of the mass catalog were generated, perturbing each mass within a gaussian of sigma equal to its error. Unless stated differently in the relevant Figure captions, we use the median values of the Monte-Carlo simulations in all figures. Error bars take into account both poissonian errors on the median counts \citep{gehr1986}, and 16-84$^{th}$ percentile values of each distribution. 

In Fig.~\ref{fig:morph_MF} we show the V/V$_{max}$ corrected stellar mass functions, split by morphological type, in three redshift bins. The redshift intervals were chosen to have comparable numbers of objects in each bin. A dotted vertical line shows the mass completeness for each redshift bin. We find that, at all redshifts, the late--type objects dominate the low--mass end of the mass function, while the early-- and intermediate--type objects dominate the high--mass end. In addition, Fig.~\ref{fig:morph_MF} shows that the relative contribution of disks to the high--mass end of the mass function increases with redshift. This suggests that: {\it i)} the morphological mix at the high--mass end evolves with redshift; {\it ii)} the transition mass increases with redshift. The latter finding was already suggested in \citet{bundy2005}, but its significance was hampered by their mass completeness limit. In our case, the transition seems to happen well above the estimated mass completeness. As Fig.~\ref{fig:morph_MF} shows, at $z\sim1$ the disks' and bulges' contributions become comparable at $\approx1\times10^{11}$M$_\odot$.
\begin{figure}
\begin{center}
\includegraphics[angle=0,width=0.85\linewidth,bb = 19 145 580 700]{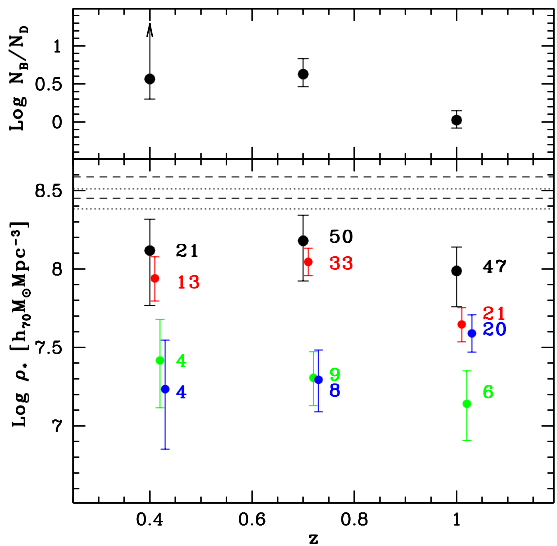}
\caption{{\bf Top panel:} The redshift evolution of the ratio of bulges and disks number densities. Error bars show the 5$^{th}$ and 95$^{th}$ percentiles of the ratio distributions. {\bf Bottom panel:} Stellar mass densities  for objects with M$_*\ge7\times10^{10}$M$_\odot$ (color coding according to galaxy morphology, as in Fig.~\ref{fig:morph_MF}). Points are slightly shifted in redshift for clarity. The median numbers of objects in each class are also labeled. Error bars account for poissonian errors, mass uncertainties, and cosmic variance (for the total values only) as described in the text. Dotted and dashed lines indicate the 3$\sigma$ range estimate obtained at redshift 0, with the same mass cut and IMF, from \citet{cole2001} and  \citet{bell2003}, respectively. \label{morph_mix_evol}}
\end{center}
\end{figure}
In order to better explore the evolution of the morphological mix at the high--mass end, we cut the three subsamples at the same common mass completeness (M$_*\ge7\times10^{10}$M$_\odot$) hereafter. This guarantees a fair comparison between the three redshift bins.
The upper panel of Fig.~\ref{morph_mix_evol} shows the behavior of the ratio of the number densities of bulges and disks. At increasing redshift, this ratio is found to systematically decrease with respect to the no--evolution hypothesis at more than 3$\sigma$ significance. In the bottom panel of Fig.~\ref{morph_mix_evol}, we show the stellar mass density split by morphological type at different redshifts. The error estimates of the total values include the cosmic variance contribution, estimated as in \citet{somerville2004}. 

In agreement with previous studies (see \S1), we find that at $z\sim1$ massive objects host almost half of the stellar mass contained in similarly massive objects at $z=0$. In addition, we find that the contribution to the total mass budget from early and late type galaxies is almost equal at $z\sim1$, but strongly evolves with redshift: there is a {\it mass pouring} from disk systems to  bulge--dominated objects. Our findings require a scenario in which massive objects almost double their mass from redshift 1 to 0, and become more and more bulge--dominated systems as time goes by. 

\begin{figure}
\begin{center}
\includegraphics[angle=0,width=0.85\linewidth,bb = 19 145 580 540]{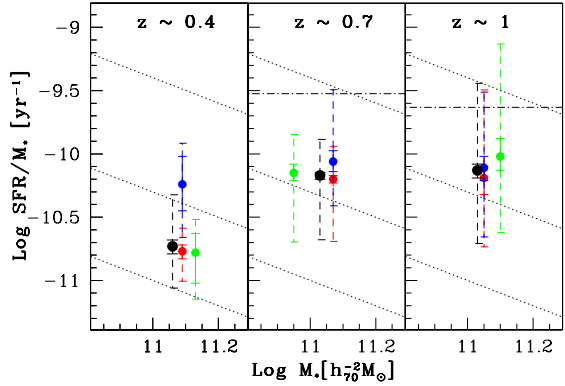}
\caption{The median SSFR for different morphological types. The coloured symbols (same colour coding as in previous figures) show the median values in both stellar mass and SSFR. Dashed error bars represent the median 10$^{th}$ and 90$^{th}$ percentiles of the Monte-Carlo realizations, while the solid error bars show 1$\sigma$ uncertainties on the median values. Tilted lines are  constant SFR values (from the bottom 1,5,40 M$_\odot yr^{-1}$) while the horizontal lines mark the doubling mass lines (only for the last two redshift bins -- see text for details). \label{ssfr}}
\end{center}
\end{figure}

A possibility is that {\it in situ} star formation increases the mass and secular evolution modifies the morphology \citep{korm2004}. Under simplified assumptions, we can estimate whether the star formation rates (SFR) of the galaxies in our three redshift subsamples can account for a doubling of the stellar mass by z = 0. Following \citet{madau1998}, the SFR of individual objects can be estimated from the restframe UV luminosity as ${\rm SFR}_{2800}=1.27\times10^{-28}\times L_{2800}$ in units of M$_\odot$yr$^{-1}$, where the constant factor is computed for a Salpeter IMF. We apply a dust attenuation correction of A$_{2800}=1$mag to the whole sample. This median value, assumed not to depend on redshift, is derived as in \citet{gabphd} by comparing the total stellar mass density and the integral of the SFR density at different look--back times. Fig.~\ref{ssfr} shows the median SSFR values for the three redshift bins and for the different morphological types, cut at the same mass completeness as described above. Since it is not possible to know the individual star formation histories, we derive the median SFR decline with cosmic time from the massive galaxies subsample, and we assume that all the objects in our sample follow this median star formation history. In this way we can estimate, at every redshift, the minimum SSFR that enables an object to double its mass by z = 0. These minimum SSFRs are shown in the plot as horizontal lines at the last two redshift bins considered. It appears that only a small fraction of the $z\sim1$ sample would be able to double its mass. 

An alternative and more likely solution \citep[see also][]{brinchmanneellis2000} is a continuous accretion of smaller mass galaxies, and possibly the merging of massive disks, which can account for both the changing total mass budget and the morphological mix evolution.

In the past a number of works \citep[e.g.][]{brinchmanneellis2000,feulnerb,feulner2005,bauer2005} have focused on the SSFR redshift evolution and consistently found an increase of the SSFRs with redshift. Hence the bulk of star formation in massive galaxies was pushed to $z\ge2$, in agreement with the downsizing scenario \citep{cowie1996}. 

Here for the first time we explore the SSFR evolution split by different morphological types. We find, for objects with M$_*\ge7\times10^{10}$M$_\odot$, a different behavior in the SSFR redshift evolution for the different morphological types. The median SSFR of late-type objects shows only a mild, if any, evolution up to redshift 1, while the median values of early and intermediate types are clearly increasing with redshift. Since these latter populations are dominanting in numbers, they drive the redshift evolution of the whole massive galaxies distribution. This finding, together with the previous arguments, suggests that the morphological evolution of massive disk galaxies toward more and more bulge dominated systems is accompanied by a decrease in the specific star formation rate of their end products. 

\citet{bell2005} show that a large part of the star formation in massive galaxies at $z\sim0.7$ is due to normal spiral galaxies. Moreover, they conclude that turning off star formation in these objects is responsible for the observed decline of cosmic star formation therafter. We find a good agreement at $z\sim0.7$ with their result but also a hint that the decline of star formation in massive galaxies is linked more to the evolution of bulge--dominated galaxies.
\section{Conclusions}
\label{sec_conclusions}
In this Letter we have studied the contribution of disks and bulges to the evolution of the stellar mass function up to $z\sim1$. 

We agree with \citet{bundy2005} that the transition mass, i.e. the mass at which disks become dominant in the relative contribution to the total mass function, increases with redshift. We estimate that at $z\sim1$ the transition mass is up to a factor 2 larger than its measured local value. 

We show that the morphological mix evolves with redshift. At $z\sim1$ early and late type galaxies contribute nearly equally to the total mass budget in massive objects (i.e. with M$_*\ge7\times10^{10}$M$_\odot$). We suggest that merging events must play a key role in the {\it mass pouring} from disks to bulges. 

We find a different behavior of the SSFR, i.e. the star formation rate per unit stellar mass, as a function of redshift for the different morphological types. The median SSFR of late-type objects shows almost no evolution up to $z\sim1$. Conversely, median SSFRs for early and intermediate types increase systematically. Since these latter morphological types are dominating in numbers, they drive the total SFR evolution of massive galaxies. 
This suggests a scenario where the morphological evolution of massive disk galaxies through merging is followed by the decrease of the star formation in their bulge--dominated descendants, maybe after a burst of star formation that exhausts the available gas. 

\acknowledgments
We wish to thank the referee for valuable comments that pushed us to strengthen our results and to improve their presentation. M.P. is glad to thank G. De Lucia, D. Pierini, G. Rudnick and V. Strazzullo for a careful reading of the manuscript and for many insightful comments and suggestions. This work was supported by the Deutsche Forschungsgemeinschaft through the SFB 375 grant.

\end{document}